\documentstyle[12pt,preprint,prl,aps,amsfonts,floats]{revtex} 
\tighten 
\begin{document}  
%
\title{Remarks on the neutrino oscillation formula}

\author{Massimo Blasone${}^{a,b}$ and Giuseppe
Vitiello${}^{b}$\thanks{e-mail: m.blasone@ic.ac.uk,
vitiello@physics.unisa.it}}

\address{${}^{a}$ Blackett Laboratory, Imperial College, Prince Consort
Road, \\ London SW7 2BZ, U.K. }  \address{${}^{b}$Dipartimento di Fisica
dell'Universit\`a   and INFN, Gruppo Collegato, Salerno \\  I-84100
Salerno, Italy} 
\maketitle 
\begin{abstract}  
We show that the neutrino oscillation formula recently derived  in the
quantum field theory framework holds true despite the arbitrariness in
the mass parameter for the flavor fields. This formula is exact and
exhibits new features with respect to the usual Pontecorvo formula, which
is however valid in the relativistic limit. 
\end{abstract}  
\vspace{0.3cm}

P.A.C.S.: 14.60.Pq, 11.15.Tk, 12.15.Ff 

$$   $$

\section{Introduction}

In view of the great theoretical and experimental interest\cite{Sol}
several papers have been recently devoted to the  quantum field
theoretical approach to the neutrino mixing and
oscillations\cite{GKL92,BV95,BHV99,FHY99}.  In particular, the study of
the 
generator of the Pontecorvo mixing transformations\cite{BP78} has
shown\cite{BV95} that the Hilbert space where the mixed (flavor) field
operators are defined is unitarily inequivalent, in the infinite volume
limit, to the Hilbert space for the original (unmixed) field
operators. Such  a finding leads to a novel understanding of the field
mixing and to a new, exact oscillation formula\cite{BHV99}, which
reduces to the Pontecorvo one in the relativistic limit. 
                                           
As a matter of fact, the problem of the definition and of the physical
interpretation of the state space for the flavor fields is a
controversial one and it is the object of still open discussions
\cite{GKL92,BV95,BHV99,FHY99}. On the other hand, the
discovery\cite{BV95} 
that the Pontecorvo field mixing transformation is a  non-unitarily
implementable transformation rests on firm mathematical grounds, so
that it cannot be ignored in any discussion on the field mixing
problems. In  a recent paper \cite{FHY99} it has been thus considered
the degree of arbitrariness involved in the construction of the flavor
states starting from the results of refs. \cite{BV95,BHV99}
(denoted by the authors of ref.\cite{FHY99} as the BV formalism). 

By using a more general mass parameterization than the one adopted in
the 
BV formalism, the conclusion of the authors of ref. \cite{FHY99} has
been that, since  the mass arbitrariness shows up in the oscillation
probability, the conclusions drawn in refs.\cite{BV95} are
unphysical. In particular, they conclude that the procedure adopted in
ref.\cite{BV95} by choosing a specific mass parameterization has no
physical basis.  They then examine the structure of the neutrino
propagator. However,  although the physical poles of the propagator are
shown to coincide with the eigenvalues  of the mass matrix in the
Lagrangian and appear not to be affected by the mass parameterization
arbitrariness, they are not able to escape the arbitrariness difficulty
in the oscillation formula. Thus they are finally forced to admit that
it is still necessary to investigate in detail how to derive
oscillations formulas reflecting real experimental situations on the
basis of the field theory. 

Motivated by such a necessity, in the present paper we consider the
arguments of ref.\cite{FHY99} and we show that, even by applying the
more general mass parameterization there adopted, 
the mass arbitrariness
disappears from the exact oscillation formula as derived in
ref.\cite{BHV99}. Actually, the oscillation formula considered in
ref.\cite{FHY99} is the approximate one which was derived in
ref.\cite{BV95}, and not its exact form obtained in the Green's
function formalism of ref.\cite{BHV99}. 

The conclusions of ref.\cite{FHY99} about the unphysical basis of the
analysis or refs.\cite{BV95,BHV99} are thus ruled out. On the
contrary, even though the problem of the interpretation of the flavor
space may still be object of discussion, the field theoretical formalism
derived in refs. \cite{BV95,BHV99} appears to be sound and,
what is most interesting, it leads to the oscillation formula which is
experimentally testable. 

In the present paper we also shortly comment on the physical meaning of
the mass arbitrariness which has been introduced by the authors of
ref. \cite{FHY99} without explicit justification. These comments are
also useful in order to clarify the physical meaning of the more
general transformations used in ref.\cite{FHY99} and of the particular
choice adopted in the BV formalism. 

The paper is organized as follows. In Section II generalities on the
formalism are reviewed, also considering the generalization of
ref.\cite{FHY99}. In Section III the exact oscillation formula is
rederived in the formalism of ref.\cite{FHY99} and it is shown to be
independent of the mass arbitrariness. Section IV  is devoted to further
comments and conclusions.

\section{Mixed fermions}

To be definite, let us consider an oversimplified model with two 
(Dirac) fermion fields with a mixed mass term:
\begin{equation}\label{lag2} 
{\cal L} = {\bar \nu}_e\left ( i\!\not\!\partial -m_{e}\right)\nu_e +
{\bar    \nu}_\mu\left(i\!\not\!\partial - m_{\mu}\right)\nu_\mu   - \; 
m_{e \mu} \;\left({\bar \nu}_{e}   \nu_{\mu} + {\bar \nu}_{\mu}
\nu_{e}\right)    
\;.\end{equation}  
The above Lagrangian is sufficient in order to describe the
single-particle evolution of a mixed state, i.e. neutrino
oscillations\cite{BHV99}, and can be fully diagonalized by the
transformation
\begin{eqnarray} \nonumber  
\nu_{e}(x)   &=&\nu_{1}(x)\,\cos\theta +    \nu_{2}(x)\,\sin\theta\\  
\label{rot1}  
\nu_{\mu}(x) &=&-\nu_{1}(x)\,\sin\theta    + \nu_{2}(x)\,\cos\theta
\;,\end{eqnarray}  
where $\theta$ is the mixing angle and   
$ m_{e} = m_{1}\cos^{2}\theta +
m_{2} \sin^{2}\theta~$,  $m_{\mu} = m_{1}\sin^{2}\theta + m_{2}
\cos^{2}\theta~$,   $m_{e\mu} =(m_{2}-m_{1})\sin\theta \cos\theta\,$.
$\nu_1$ and $\nu_2$ therefore   are non-interacting, free fields,
anticommuting with each   other at any space-time point.   They are
explicitly given by  
\begin{equation}\label{2.2}\nu_{i}(x) =   
\frac{1}{\sqrt{V}} \sum_{{\bf k},r} \left[u^{r}_{{\bf k},i}
\alpha^{r}_{{\bf k},i}(t)\:+    v^{r}_{-{\bf k},i}\beta^{r\dag }_{-{\bf
k},i}(t)   \right] e^{i {\bf k}\cdot{\bf x}} , \; \;
\; ~  i=1,2 \;.   
\end{equation}  
with $\alpha^{r}_{{\bf k},i}(t)=e^{-i\omega_{i} t}\alpha^{r}_{{\bf
k},i}(0)$, $\beta^{r}_{{\bf k},i}(t)=e^{-i\omega_{i} t}\beta^{r}_{{\bf
k},i}(0)$ and  $\omega_{i}=\sqrt{{\bf k}^2+m_i^2}$.  Here and in the
following we use $t\equiv x_0$, when no misunderstanding  arises.   The
vacuum for the $\alpha_i$ and $\beta_i$ operators is denoted by
$|0\rangle_{1,2}$:  $\; \; \alpha^{r}_{{\bf k},i}|0\rangle_{12}=
\beta^{r }_{{\bf    k},i}|0\rangle_{12}=0$.   The anticommutation
relations are the usual ones (see ref.\cite{BV95}).  The orthonormality
and  completeness relations are:
\begin{equation} 
u^{r\dag}_{{\bf k},i} u^{s}_{{\bf k},i} =   v^{r\dag}_{{\bf k},i}
v^{s}_{{\bf k},i} = \delta_{rs}  \;,\quad  u^{r\dag}_{{\bf k},i}
v^{s}_{-{\bf k},i} =  v^{r\dag}_{-{\bf k},i} u^{s}_{{\bf k},i} =
0\;,\quad \sum_{r}(u^{r}_{{\bf k},i} u^{r\dag}_{{\bf k},i} +
v^{r}_{-{\bf k},i} v^{r\dag}_{-{\bf k},i}) = I \;. 
\end{equation}

The fields $\nu_e$ and $\nu_\mu$ are thus    completely determined
through    eq.(\ref{rot1}), which can be rewritten in the following
form (we use $(\sigma,j)=(e,1) , (\mu,2)$):
\begin{eqnarray}\label{exnue1}  
\nu_{\sigma}(x)     &=& G^{-1}_{\theta}(t)\, \nu_{j}(x)\, G_{\theta}(t)
= \frac{1}{\sqrt{V}} \sum_{{\bf k},r}   \left[ u^{r}_{{\bf k},j}
\alpha^{r}_{{\bf k},\sigma}(t) +    v^{r}_{-{\bf k},j}
\beta^{r\dag}_{-{\bf k},\sigma}(t) \right]  e^{i {\bf k}\cdot{\bf x}}, 
\\ \label{BVgen} 
G_{\theta}(t) &=& \exp\left[\theta \int d^{3}{\bf x}
\left(\nu_{1}^{\dag}(x)\nu_{2}(x) -    \nu_{2}^{\dag}(x)
\nu_{1}(x)\right)\right]
\end{eqnarray}
where $G_{\theta}(t)$    is the generator of the mixing transformations
(\ref{rot1}) (see  ref.\cite{BV95} for a discussion of its properties). 

Eq.(\ref{exnue1}) gives an expansion of the flavor fields $\nu_{e}$ and
$\nu_{\mu}$ in the same basis of $\nu_{1}$ and $\nu_{2}$.  In the BV
formalism the flavor annihilation operators are then identified with
\begin{equation}\label{BVoper}
\left(\begin{array}{c} \alpha^{r}_{{\bf k},\sigma}(t)\\
\beta^{r\dag}_{{-\bf k},\sigma}(t)
\end{array}\right)
= G^{-1}_{\theta}(t)  \left(\begin{array}{c} \alpha^{r}_{{\bf k},j}(t)\\
\beta^{r\dag}_{{-\bf k},j}(t)
\end{array}\right)
G_{\theta}(t)
\end{equation}  
The BV flavor vacuum is defined as $|0(t)\rangle_{e,\mu}\equiv
G^{-1}_{\theta}(t)|0\rangle_{1,2}\, $. 

The explicit expression of the flavor annihilation operators is (we
choose ${\bf k}=(0,0,|{\bf k}|)$): 
\begin{equation}\label{BVmatrix}  
\left(\begin{array}{c} {\alpha}^{r}_{{\bf k},e}(t)\\ {\alpha}^{r}_{{\bf
k},\mu}(t)\\ {\beta}^{r\dag}_{{-\bf k},e}(t)\\ {\beta}^{r\dag}_{{-\bf
k},\mu}(t)
\end{array}\right)
\, = \, \left(\begin{array}{cccc} c_\theta &  s_\theta\, |U_{{\bf k}}|
&0 & s_\theta \, \epsilon^{r} \,|V_{{\bf k}}| \\ - s_\theta\, |U_{{\bf
k}}|& c_\theta 
&  s_\theta \,\epsilon^{r} \,|V_{{\bf k}}| & 0  \\ 0& - s_\theta
\,\epsilon^{r} \,|V_{{\bf k}}| 
&c_\theta & s_\theta \,|U_{{\bf k}}| \\ - s_\theta \,\epsilon^{r}
\,|V_{{\bf k}}| & 0 & 
- s_\theta\, |U_{{\bf k}}| & c_\theta \\
\end{array}\right)
\left(\begin{array}{c} \alpha^{r}_{{\bf k},1}(t)\\ \alpha^{r}_{{\bf
k},2}(t)\\ \beta^{r\dag}_{{-\bf k},1}(t)\\ \beta^{r\dag}_{{-\bf k},2}(t)
\end{array}\right)
\end{equation}
where $c_\theta\equiv \cos\theta$, $s_\theta\equiv \sin\theta$,
$\epsilon^r\equiv (-1)^r$, and 
\begin{eqnarray}
&&|U_{{\bf k}}|\equiv u^{r\dag}_{{\bf k},2}u^{r}_{{\bf k},1}=
v^{r\dag}_{-{\bf k},1}v^{r}_{-{\bf k},2} \\
\label{2.37}
&&|V_{{\bf k}}|\equiv \epsilon^{r}\;u^{r\dag}_{{\bf  k},1}v^{r}_{-{\bf
k},2}= -\epsilon^{r}\;u^{r\dag}_{{\bf k},2}v^{r}_{-{\bf k},1}\,. 
\end{eqnarray}
We have:
\begin{eqnarray}
&&|U_{{\bf
k}}|=\left(\frac{\omega_{k,1}+m_{1}}{2\omega_{k,1}}\right)^{\frac{1}{2}}
\left(\frac{\omega_{k,2}+m_{2}}{2\omega_{k,2}}\right)^{\frac{1}{2}}
\left(1+\frac{|{\bf k}|^{2}}{(\omega_{k,1}+m_{1})
(\omega_{k,2}+m_{2})}\right)\\
\label{2.39}
&&|V_{{\bf k}}|=\left(\frac{\omega_{k,1}+m_{1}}{2\omega_{k,1}}
\right)^{\frac{1}{2}}
\left(\frac{\omega_{k,2}+m_{2}}{2\omega_{k,2}}\right)^{\frac{1}{2}}
\left(\frac{|{\bf k}|}{(\omega_{k,2}+m_{2})}-
\frac{|{\bf k}|}{(\omega_{k,1}+m_{1})}\right) 
\end{eqnarray}
\begin{equation}\label{2.40}
|U_{{\bf k}}|^{2}+|V_{{\bf k}}|^{2}=1 
\end{equation}

\medskip

It has been recently noticed\cite{FHY99}, however, that  expanding the
flavor fields in the same basis as the (free) fields with definite
masses is actually a special choice, and that a more general possibility
exists. 

In other words, in the expansion eq.(\ref{exnue1}) one could  use
eigenfunctions with arbitrary masses $\mu_\sigma$, and therefore not
necessarily the same as the masses which appear in the Lagrangian.  On
this basis, the authors of ref.\cite{FHY99} have generalized the BV
transformation (\ref{BVoper}) by writing the flavor fields as
\begin{eqnarray}\label{exnuf2}  
\nu_{\sigma}(x)     &=& \frac{1}{\sqrt{V}} \sum_{{\bf k},r}   \left[
u^{r}_{{\bf k},\sigma} {\widetilde \alpha}^{r}_{{\bf k},\sigma}(t) +
v^{r}_{-{\bf k},\sigma} {\widetilde \beta}^{r\dag}_{-{\bf k},\sigma}(t)
\right]  e^{i {\bf k}\cdot{\bf x}} ,
\end{eqnarray}
where $u_{\sigma}$ and $v_{\sigma}$ are the helicity eigenfunctions with
mass $\mu_\sigma$\footnote{The use of such a basis simplifies
considerably calculations with respect to the original choice of
ref.\cite{BV95}.}. We denote by a tilde the generalized flavor operators
introduced in ref.\cite{FHY99} in order to distinguish them from the
ones defined in eq.(\ref{BVoper}).  The expansion eq.(\ref{exnuf2}) is
more general than the one in eq.(\ref{exnue1}) since the latter
corresponds to the particular choice $\mu_e\equiv m_1$, $\mu_\mu \equiv
m_2$. 
 
The relation between the flavor and the mass operators is now:
\begin{equation}\label{FHYoper}  
\left(\begin{array}{c} {\widetilde \alpha}^{r}_{{\bf k},\sigma}(t)\\
{\widetilde \beta}^{r\dag}_{{-\bf k},\sigma}(t)
\end{array}\right)
= K^{-1}_{\theta,\mu}(t)  \left(\begin{array}{c} \alpha^{r}_{{\bf
k},j}(t)\\ \beta^{r\dag}_{{-\bf k},j}(t)
\end{array}\right)
K_{\theta,\mu}(t) ~~,
\end{equation}
with $(\sigma,j)=(e,1) , (\mu,2)$ and 
where $K_{\theta,\mu}(t)$ is the generator of the transformation
(\ref{rot1}) and can be written as
\begin{eqnarray}\label{FHYgen}
K_{\theta,\mu}(t)&=& I_{\mu}(t)\, G_{\theta}(t) \\ \label{Igen}
I_{\mu}(t)&=& \prod_{{\bf k}, r}\, \exp\left\{ i
\mathop{\sum_{(\sigma,j)}} \xi_{\sigma,j}^{\bf k}\left[
\alpha^{r\dag}_{{\bf k},j}(t)\beta^{r\dag}_{{-\bf k},j}(t) +
\beta^{r}_{{-\bf k},j}(t)\alpha^{r}_{{\bf k},j}(t) \right]\right\}
\end{eqnarray}
with $\xi_{\sigma,j}^{\bf k}\equiv (\chi_\sigma - \chi_j)/2$ and
$\cot\chi_\sigma = |{\bf k}|/\mu_\sigma$, $\cot\chi_j = |{\bf k}|/m_j$.
For $\mu_e\equiv m_1$, $\mu_\mu \equiv m_2$ one has $I_{\mu}(t)=1$. 

The explicit  matrix form of the flavor operators is\cite{FHY99}:
\begin{equation}\label{FHYmatrix}  
\left(\begin{array}{c} {\widetilde \alpha}^{r}_{{\bf k},e}(t)\\
{\widetilde \alpha}^{r}_{{\bf k},\mu}(t)\\ {\widetilde
\beta}^{r\dag}_{{-\bf k},e}(t)\\ {\widetilde \beta}^{r\dag}_{{-\bf
k},\mu}(t)
\end{array}\right)
\, = \, \left(\begin{array}{cccc} c_\theta\, \rho^{\bf k}_{e1} &
s_\theta \,\rho^{\bf k}_{e2}  &i c_\theta \,\lambda^{\bf k}_{e1} & i
s_\theta \,\lambda^{\bf k}_{e2}  \\ - s_\theta \,\rho^{\bf k}_{\mu 1} &
c_\theta \,\rho^{\bf k}_{\mu 2} &- i s_\theta \,\lambda^{\bf k}_{\mu 1}
& i c_\theta \,\lambda^{\bf k}_{\mu 2}  \\ i c_\theta \,\lambda^{\bf
k}_{e1} & i s_\theta \,\lambda^{\bf k}_{e2} &c_\theta\, \rho^{\bf
k}_{e1} & s_\theta \,\rho^{\bf k}_{e2}  \\ - i s_\theta \,\lambda^{\bf
k}_{\mu 1} & i c_\theta\, \lambda^{\bf k}_{\mu 2} &- s_\theta\,
\rho^{\bf k}_{\mu 1} & c_\theta\, \rho^{\bf k}_{\mu 2} \\
\end{array}\right)
\left(\begin{array}{c} \alpha^{r}_{{\bf k},1}(t)\\ \alpha^{r}_{{\bf
k},2}(t)\\ \beta^{r\dag}_{{-\bf k},1}(t)\\ \beta^{r\dag}_{{-\bf k},2}(t)
\end{array}\right)
\end{equation}
where $c_\theta\equiv \cos\theta$, $s_\theta\equiv \sin\theta$ and
\begin{eqnarray}\label{rho}
 \rho^{\bf k}_{a b} \delta_{rs}&\equiv& \cos\frac{\chi_a - \chi_b}{2}
\delta_{rs}= u^{r\dag}_{{\bf k},a} u^{s}_{{\bf k},b} =
v^{r\dag}_{-{\bf k},a} v^{s}_{-{\bf k},b} \\ \label{lambda} i
\lambda^{\bf k}_{a b}\delta_{rs} &\equiv& i \sin\frac{\chi_a -
\chi_b}{2} \delta_{rs} = u^{r\dag}_{{\bf k},a} v^{s}_{-{\bf k},b} =
v^{r\dag}_{-{\bf k},a} u^{s}_{{\bf k},b}
\end{eqnarray}
with $a,b = 1,2,e,\mu$. Since $\rho^{\bf k}_{1 2}=|U_{{\bf k}}|$ and
$i \lambda^{\bf k}_{1 2}=\epsilon^r |V_{{\bf k}}|$, etc., 
the operators (\ref{FHYmatrix})
reduce to the ones in eqs.(\ref{BVmatrix}) when $\mu_e\equiv m_1$ and
$\mu_\mu \equiv m_2$\footnote{In performing such an identification, one
should take into account that the operators for antiparticles differ for
a minus sign, related to the different spinor bases used in the
expansions (\ref{exnue1}) and (\ref{exnuf2}). Such a sign difference 
is however irrelevant in what follows.}.

The generalization of the BV flavor vacuum, which is
annihilated by the flavor operators given by eq.(\ref{FHYoper}), is now
written as\cite{FHY99}:
\begin{equation}\label{FHYvac}
|{\widetilde 0(t)}\rangle_{e,\mu}\equiv
K^{-1}_{\theta,\mu}(t)|0\rangle_{1,2} ~~. 
\end{equation}
For $\mu_e\equiv m_1$ and $\mu_\mu \equiv m_2$, this state reduces to
the BV flavor vacuum $| 0(t)\rangle_{e,\mu}$ above defined. 


For the  considerations which follow, it is also useful to report here
the relation, given in ref.\cite{FHY99}, between the general flavor
operators of eq.(\ref{FHYoper}) and the BV ones:
\begin{eqnarray}\label{FHYBVa}  
\left(\begin{array}{c} {\widetilde \alpha}^{r}_{{\bf k},\sigma}(t)\\
{\widetilde \beta}^{r\dag}_{{-\bf k},\sigma}(t)
\end{array}\right)
&=& J^{-1}_{\mu}(t)  \left(\begin{array}{c} \alpha^{r}_{{\bf
k},\sigma}(t)\\ \beta^{r\dag}_{{-\bf k},\sigma}(t)
\end{array}\right)J_{\mu}(t) ~~,
\\ \label{FHYBVb}
J_{\mu}(t)&=& \prod_{{\bf k}, r}\, \exp\left\{ i
\mathop{\sum_{(\sigma,j)}} \xi_{\sigma,j}^{\bf k}\left[
\alpha^{r\dag}_{{\bf k},\sigma}(t)\beta^{r\dag}_{{-\bf k},\sigma}(t) +
\beta^{r}_{{-\bf k},\sigma}(t)\alpha^{r}_{{\bf k},\sigma}(t) 
\right]\right\}\,.
\end{eqnarray}

\section{The oscillation formula}

In the formal framework of the previous Section the annihilation
operators and the vacuum for mixed fermions are defined
self-consistently, and the Hilbert space for mixed neutrinos can thus be
constructed. Such an Hilbert space, however, has built in the
arbitrariness related with the mass parameters $\mu_{\sigma}$, $\sigma
= e, \mu$. According to the authors of ref.\cite{FHY99}, such an
arbitrariness also shows up in the final expression of the oscillation
probability, which is a non-acceptable result since the theory arbitrary
parameters should have no effects on observable quantities.  The full
construction, although mathematically consistent, would then be
questionable from a physical point of view. Now we show that the
analysis of ref.\cite{FHY99} is not complete, that the exact
oscillation probabilities are independent of the arbitrary mass
parameters 
and therefore the conclusion of ref.\cite{FHY99} is ruled out. 

The main point is that the authors of ref.\cite{FHY99} miss to compute
the full oscillation probability  whose exact form is presented in
ref.\cite{BHV99} and it is there obtained in the Green's function
formalism. In fact, the statement of ref.\cite{FHY99} that the
oscillation formula ``seems  not to be correct'' since it is based on the
one neutrino state, which does depend on $\mu_e$ and $\mu_\mu$, is not
correct: as we show below, it is possible to calculate the oscillation
probabilities by using the arbitrary mass formalism of Section II, getting
a result which is independent of the arbitrary masses $\mu_e$ and
$\mu_\mu$ and coincides with the one  of ref.\cite{BHV99}. 

In the line of ref.\cite{BHV99}, let us consider the propagator for the
flavor fields, which has to be defined on the proper (flavor)
vacuum. Notice that here we perform the computations in the generalized
BV formalism of ref.\cite{FHY99}. The propagators are then given by:
\begin{equation}\label{matrix2}
\left(\begin{array}{cc}   {\widetilde G}_{ee}(x,y)& {\widetilde G}_{\mu
e}(x,y)  
\vspace{0.1cm}\\  {\widetilde G}_{e \mu}(x,y)& {\widetilde G}_{\mu
\mu}(x,y)  
\end{array} \right)  
\equiv  \,_{e,\mu}\langle {\widetilde 0}(y_0) | \!
\left(\begin{array}{cc}   T \left[\nu_{e}(x) \bar{\nu}_{e}(y)\right]&
T \left[\nu_{\mu}(x) \bar{\nu}_{e}(y)\right]  
\vspace{0.1cm} \\  T \left[\nu_{e}(x) \bar{\nu}_{\mu}(y)\right]&   T
\left[\nu_{\mu}(x) \bar{\nu}_{\mu}(y)\right]  
\end{array} \right)\!|{\widetilde 0}(y_0) \rangle_{e,\mu}  \, ,
\end{equation}
where the state used is the one defined in eq.(\ref{FHYvac}). These
propagators clearly {\em do depend} on the arbitrary parameters 
$\mu_e$ and
$\mu_\mu$, which are present in $|{\widetilde 0}(y_0) \rangle_{e,\mu} $.
However, the propagator is not a measurable quantity: on the  contrary,
the oscillation probability, which can be defined in terms of it, is 
measurable and should not be affected by any arbitrary
parameters. 

Let us then consider the case of an initial electron neutrino which
evolves (oscillates) in time. The two relevant propagators
are\cite{BHV99}:
\begin{eqnarray}\label{propee}
i {\widetilde G}^{>}_{ee}(t,{\bf x};0,{\bf y}) =    {}_{e,\mu}\langle
{\widetilde 0}|\nu_{e}(t,{\bf x}) \;   \bar{\nu}_{e}(0,{\bf y})
|{\widetilde 0}\rangle_{e,\mu} 
\\ \label{propem} 
i {\widetilde
G}^{>}_{\mu e} (t,{\bf x};0,{\bf y}) =    {}_{e,\mu}\langle {\widetilde
0}|    \nu_{\mu}(t,{\bf x}) \; \bar{\nu}_{e}(0,{\bf y})   |{\widetilde
0}\rangle_{e,\mu}
\end{eqnarray}  
where $|{\widetilde 0}\rangle_{e,\mu} \equiv |{\widetilde 0}(t=0)
\rangle_{e,\mu} $. 
As discussed in \cite{BHV99}, there are four distinct transition
amplitudes which can be defined from  the above propagators. In the
present generalized case, we have to use the wave functions $u_{\sigma}$
and $v_{\sigma}$,  instead of $u_{j}$ and $v_{j}$ which were used in
\cite{BHV99}. This is in line with the discussion of the previous
Section: the choice of the basis in which we expand the flavor fields
determines the relevant annihilators and then the vacuum. 

The amplitudes are then obtained as\footnote{With respect to the ones
defined in ref.\cite{BHV99}, we omit here an (irrelevant) phase
factor. This is due to the different definition of the flavor operators
 -- see eq.(\ref{BVmatrix}).}
\begin{eqnarray} \label{pee1}
{\widetilde {\cal P}}^r_{ee} ({\bf  k},t)   &\equiv& i \,u^{r
 \dag}_{{\bf k},e}\,   {\widetilde G}^>_{ee}({\bf  k},t)\,\gamma^0
 u^r_{{\bf k},e} =    \left\{{\widetilde \alpha}^r_{{\bf k},e}(t),
 {\widetilde \alpha}^{r\dag}_{{\bf k},e}(0) \right\}   
\\ \nonumber  \\ \label{pee2}   
{\widetilde {\cal P}}^r_{\bar{e}e} ({\bf  k},t)
 &\equiv&i \,v^{r \dag}_{-{\bf k},e}\,   {\widetilde G}^>_{ee}({\bf
 k},t)\,\gamma^0 u^r_{{\bf k},e} =   \left\{{\widetilde
 \beta}^{r\dag}_{-{\bf k},e}(t), {\widetilde \alpha}^{r\dag}_{{\bf
 k},e}(0) \right\}   
\\ \nonumber   \\ \label{pee3}   
{\widetilde {\cal
 P}}^r_{\mu e}({\bf  k},t)&\equiv& i \,u^{r    \dag}_{{\bf k},\mu}\,
 {\widetilde G}^>_{\mu e}({\bf  k},t)\,\gamma^0 u^r_{{\bf k},e} =
 \left\{{\widetilde \alpha}^r_{{\bf k},\mu}(t),  {\widetilde
 \alpha}^{r\dag}_{{\bf k},e}(0) \right\}    
\\ \nonumber   \\ \label{pee4}   
{\widetilde {\cal P}}^r_{\bar{\mu} e}({\bf
 k},t)&\equiv&   i \,v^{r \dag}_{-{\bf k},\mu}\,   {\widetilde G}^>_{\mu
 e}({\bf  k},t)\,\gamma^0 u^r_{{\bf k},e} =    \left\{{\widetilde
 \beta}^{r\dag}_{-{\bf k},\mu}(t), {\widetilde \alpha}^{r\dag}_{{\bf
 k},e}(0) \right\}   
\end{eqnarray}  
The explicit form of these amplitudes is rather complicated. Notice that
 all of them involve the arbitrary parameters $\mu_\sigma$. However, it
 can be verified that the following sum rule for the squared moduli is
 still valid:
\begin{equation}\label{cons}  
\left|{\widetilde {\cal P}}^r_{ee}({\bf  k},t)\right|^2 +
\left|{\widetilde {\cal P}}^r_{\bar{e}e}({\bf  k},t)\right|^2 +
\left|{\widetilde {\cal P}}^r_{\mu e}({\bf  k},t)\right|^2 +
\left|{\widetilde {\cal P}}^r_{\bar{\mu}e}({\bf  k},t)\right|^2 =1
\,,\end{equation}  

Moreover, through somewhat long direct calculation or by employing the
linear relation eq.(\ref{FHYBVa}) (cf. also the second relation in
(2.31) of ref.\cite{FHY99}) 
as well as the charge conjugation
relation between $\alpha^{r}_{{\bf k},\sigma}$ and $\beta^{r}_{-{\bf
k},\sigma}$, we obtain
\begin{equation}\label{miracle}   
\left|\left  \{{\widetilde \alpha}^{r}_{{\bf k},e}(t),   {\widetilde
\alpha}^{r \dag}_{{\bf k},e}(0) \right\}\right|^{2} \,+
\,\left|\left\{{\widetilde \beta}_{{-\bf k},e}^{r \dag}(t),  {\widetilde
\alpha}^{r \dag}_{{\bf k},e}(0) \right\}\right|^{2} \, =\,  \left|\left
\{\alpha^{r}_{{\bf k},e}(t),   \alpha^{r \dag}_{{\bf k},e}(0)
\right\}\right|^{2} \,+   \,\left|\left\{\beta_{{-\bf k},e}^{r \dag}(t),
\alpha^{r \dag}_{{\bf k},e}(0) \right\}\right|^{2}  
\end{equation}  
\begin{equation}\label{miracle2}   
\left|\left  \{{\widetilde \alpha}^{r}_{{\bf k},\mu}(t),   {\widetilde
\alpha}^{r \dag}_{{\bf k},e}(0) \right\}\right|^{2} \,+
\,\left|\left\{{\widetilde \beta}_{{-\bf k},\mu}^{r \dag}(t),
{\widetilde \alpha}^{r \dag}_{{\bf k},e}(0) \right\}\right|^{2} \, =\,
\left|\left  \{\alpha^{r}_{{\bf k},\mu}(t),   \alpha^{r \dag}_{{\bf
k},e}(0) \right\}\right|^{2} \,+   \,\left|\left\{\beta_{{-\bf
k},\mu}^{r \dag}(t),  \alpha^{r \dag}_{{\bf k},e}(0) \right\}\right|^{2}
\end{equation}  
which is the announced result. In fact in ref.\cite{BHV99} the
probabilities for oscillating neutrinos were found to be
\begin{eqnarray} \label{enumber}  
P_{\nu_e\rightarrow\nu_e}({\bf k},t)&=& \left|\left  \{\alpha^{r}_{{\bf
k},e}(t),   \alpha^{r \dag}_{{\bf k},e}(0) \right\}\right|^{2} \;+
\;\left|\left\{\beta_{{-\bf k},e}^{r \dag}(t),  \alpha^{r \dag}_{{\bf
k},e}(0) \right\}\right|^{2}   
\\ \nonumber  &=& 
1 - \sin^{2}( 2 \theta)
\left[ |U_{{\bf k}}|^{2} \;    \sin^{2} \left( \frac{\omega_{k,2} -
\omega_{k,1}}{2} t \right)   +|V_{{\bf k}}|^{2} \;   \sin^{2} \left(
\frac{\omega_{k,2} + \omega_{k,1}}{2} t \right) \right] \, ,  
\\ \nonumber {}  
\\ \label{munumber}   
P_{\nu_e\rightarrow\nu_\mu}({\bf
k},t)&=&   \left|\left\{\alpha^{r}_{{\bf k},\mu}(t),  \alpha^{r
\dag}_{{\bf k},e}(0) \right\}\right|^{2} \;+
\;\left|\left\{\beta_{{-\bf k},\mu}^{r \dag}(t),  \alpha^{r \dag}_{{\bf
k},e}(0) \right\}   \right|^{2}   
\\ \nonumber  &=&  
\sin^{2}( 2
\theta)\left[ |U_{{\bf k}}|^{2}  \;  \sin^{2} \left( \frac{\omega_{k,2}
- \omega_{k,1}}{2} t \right)    +|V_{{\bf k}}|^{2} \;   \sin^{2} \left(
\frac{\omega_{k,2} + \omega_{k,1}}{2} t \right) \right] \, . 
\end{eqnarray}  
Thus, we have proven that the generalized formalism of ref.\cite{FHY99}
leads to the same exact result of ref.\cite{BHV99}. The above formula is
 still valid in the more general case: the probabilities for oscillating
 neutrinos do not depend on any arbitrary mass parameters.

\section{Comments and Conclusion}

Although already discussed in detail in ref.\cite{BHV99}, some comments
about the oscillation formulas (\ref{enumber}),(\ref{munumber}) may be
useful, in order to better clarify why the cancellation of the unphysical
parameters occurs in eqs.(\ref{miracle}),(\ref{miracle2}). 
In order to make clearer the physical reasoning of adopting the equal time
vacua in the computation of the amplitudes, we remark
that the quantities in
eqs.(\ref{enumber}),(\ref{munumber}) are nothing but the expectation
values (on the electron neutrino state at time $t$) 
of the charge operators $Q_{\sigma}\equiv
\alpha_{\sigma}^{\dag}\alpha_{\sigma} -
\beta^{\dag}_{\sigma}\beta_{\sigma}\,$ 
($\sigma=e, \mu$ and we have
suppressed the momentum and spin indices for simplicity).  
We  have indeed  
\begin{eqnarray} \label{charge1} 
&&{\cal Q}_\sigma (t)\equiv 
\langle \nu_e(t)|Q_\sigma| \nu_e(t)\rangle\, = \, 
\left|\left\{\alpha_{\sigma}(t), \alpha^{\dag}_{e}(0) 
\right\}\right|^{2}  
\;+ \;\left|\left\{\beta_{\sigma}^{\dag}(t),
\alpha^{\dag}_{e}(0) \right\}\right|^{2}\, , 
\\[2mm] \label{charge2} 
&&\;_{e,\mu}\langle 0(t)|Q_\sigma| 0(t)\rangle_{e,\mu}\, = \, 0 \,
\quad, \quad \langle \nu_e(t)|\left(Q_e\;+\;Q_{\mu}\right)| \nu_e(t)\rangle 
\,= \,1 \, ,
\end{eqnarray}  
where  $| \nu_e(t)\rangle
\equiv \exp\left[ -i H t\right]\,\alpha^{\dag}_e | 0\rangle_{e,\mu}$.

In this way, the physically obvious fact is confirmed, that the measure
of the flavor oscillation probabilities at time $t$
(eqs.(\ref{enumber}),(\ref{munumber})) is given by the expectation value
of the flavor charges, ${\cal Q}_\sigma (t)$. On the other hand, the
already established result of ref.\cite{BHV99}, by which the Green's
functions eq.(\ref{matrix2}) are well defined because of the use of the
equal time vacua, also confirms the above physical picture from a more
formal point of view, and it is strictly related to it. 
It has been
shown in ref.\cite{BHV99} that quantities like
$_{e\mu}\langle 0(x^{0})|\nu_{\sigma}(x)
\bar \nu_{e}(0,{\vec y})|0(0)\rangle_{e\mu}$ simply vanish (in the
infinite volume limit), due to  
the unitary inequivalence of flavor vacua at different times 
(we also notice that at finite volume such a quantity does
depend on the arbitrary parameters above introduced). 

We further observe that
eq.(\ref{charge2}) simply and consistently 
expresses the conservation of the total charge. 
It is remarkable that, according to the 
analysis performed in ref.\cite{BV95},
the operator for the total charge $Q_e\,+\,Q_{\mu}$
is the Casimir operator for the $su(2)$
algebra associated with the mixing transformations eq.(\ref{rot1}), and 
consequently  it commutes with the mixing generator (\ref{BVgen}) (and
(\ref{FHYgen})). 

Finally, for the full understanding of the result
(\ref{miracle}),(\ref{miracle2}), it is essential to remark that
the charge operators $Q_\sigma$ are {\em invariant} 
under the action of the
Bogoliubov generator eq.(\ref{FHYBVb}), i.e. 
${\widetilde Q}_\sigma = Q_\sigma$, where ${\widetilde Q}_{\sigma}\equiv
{\widetilde \alpha}_{\sigma}^{\dag}{\widetilde \alpha}_{\sigma} -
{\widetilde \beta}^{\dag}_{\sigma}{\widetilde \beta}_{\sigma}$.
Besides the direct computations leading to
eqs.(\ref{miracle}),(\ref{miracle2}), such an invariance, together with
eq.(\ref{charge1}), provides a strong and immediate proof of the
independence of the oscillation formula from the $\mu_\sigma$
parameters. 

In ref. \cite{FHY99} the flavor field wave functions $u_{\sigma}$ and
$v_{\sigma}$ have been introduced which satisfy the free Dirac equations
with arbitrary mass $\mu_{\sigma}$ ($\sigma = e, \mu$).
 Its introduction has not been justified in
ref.\cite{FHY99}. Therefore a short comment about the physical meaning
of such a procedure may be in order and it  can be also 
useful for a better
understanding of the formalism. Use of the wave functions  $u_{\sigma}$ and
$v_{\sigma}$  clearly
represents a more general choice than the one made in the BV
formalism\cite{BV95}, where $\mu_e\equiv m_1$, $\mu_\mu \equiv m_2$ has 
been used. 

We observe that the mass parameter $\mu_{\sigma}$ represents the
``bare'' mass of the corresponding field and therefore it can be given any
arbitrary value.  Moreover, for $\theta = 0$ the transformation
(\ref{FHYoper}) reduces to the transformation generated by $I_{\mu}(t)$
given by eq.(\ref{Igen}): now note that  this is nothing but a
Bogoliubov transformation which, at {\it $\theta = 0$}, relates unmixed
field operators, $\alpha_j$ and, say, $a_{j}(\xi_{\sigma,j})$, of
masses $m_j$ and $\mu_{\sigma}$, respectively. In the language of the 
LSZ formalism of QFT\cite{UMT,IZ}, the $\alpha_j$ refer to
physical (free) fields and the $a_{j}(\xi_{\sigma,j})$ to Heisenberg
(interacting) fields. In the infinite volume limit, the Hilbert spaces
where the operators $\alpha_j$ and $a_j$ are respectively defined, turn
out to be unitarily inequivalent spaces. Moreover, the transformation
parameter $\xi_{\sigma,j}$ acts as a label specifying Hilbert spaces
unitarily inequivalent among  themselves (for  each (different) value of
the $\mu_{\sigma}$ mass parameter). The crucial point is that the
physically relevant space is the one associated with the observable
physical mass $m_j$, the other ones being associated with the bare
masses $\mu_{\sigma}$. It can be shown\cite{UMT} that the masses
$\mu_{\sigma}$ dynamically acquire a convenient mass shift term such
that the asymptotic physical $\alpha_j$-fields are associated with the
physical mass $m_j$ and the arbitrariness intrinsic to the bare mass
$\mu_{\sigma}$ does not affect the observables. 

Therefore, in principle any one of the  $\xi$-parameterized Hilbert
spaces can be chosen to work with (in other words, the bare masses can
be given any arbitrary value). Since, however, one is interested in
observable quantities, in the LSZ formalism\cite{UMT,IZ} 
the space one chooses to work with is the
free physical field space (associated to the $\alpha_j$ operator fields,
in our case). This is the ``particular'' choice made in the BV
formalism. In the generalized BV formalism instead, by means of the
Bogoliubov transformation explicitly given by eq.(\ref{FHYmatrix})
written for $\theta = 0$, one first moves to the operators
$a_{j}(\xi_{\sigma,j})$, leaving the $\xi$ value unspecified (i.e. for
arbitrary mass parameter $\mu_{\sigma}$) and then one considers the
mixing problem. Of course, at the end of the computations observable
quantities should not depend on the arbitrary parameters, as indeed in
this paper we have proven it happens to be.     

Here we are not going to give more details on the multiplicity of
Hilbert spaces associated with arbitrary bare mass parameters.  However,
the above comment sheds some light on the physical meaning of the
particular choice made in refs.\cite{BV95,BHV99}, and it also
suggests to us why the result of the computations presented in the
present paper actually was to be expected on a physical ground, besides
being supported by straightforward mathematics. 

\section*{Acknowledgements}

This work was partially supported by MURST, INFN, and ESF. 
 

\end{document}